\documentclass[pra,aps,twocolumn,showpacs,superscriptaddress,nofootinbib,floatfix]{revtex4-1}
\usepackage[colorlinks=true,citecolor=blue,linkcolor=magenta]{hyperref}
\usepackage{graphicx,stackrel}
\usepackage{dcolumn}
\usepackage{bm}
\usepackage[bbgreekl]{mathbbol}
\usepackage{amsmath,dsfont}
\usepackage[normalem]{ulem}
\usepackage{color}
\usepackage[ruled,vlined]{algorithm2e}
\usepackage[utf8]{inputenc}

\usepackage{booktabs}

\newcommand{\bigO}{\mathcal{O}}

\begin{document}
\title{Embedding the MIS problem for non-local graphs with bounded degree using 3D arrays of atoms}
\author{Constantin Dalyac}
\affiliation{Pasqal, 2 avenue Augustin Fresnel, 91120 Palaiseau}
\affiliation{LIP6, CNRS, Sorbonne Université, 4 Place Jussieu, 75005 Paris, France}

\author{Loïc Henriet}
\affiliation{Pasqal, 2 avenue Augustin Fresnel, 91120 Palaiseau}

\date{\today}
\begin{abstract}
In the past years, many quantum algorithms have been proposed to tackle hard combinatorial problems. These algorithms, which have been studied in depth in complexity theory, are at the heart of many industrial applications. In particular, the Maximum Independent Set (MIS) is a known NP-hard problem that can be naturally encoded in Rydberg atom arrays. By representing a graph with an ensemble of neutral atoms one can leverage Rydberg dynamics to naturally encode the constraints and the solution to MIS. However, the classes of graphs that can be directly mapped ``node-to-atom" on such devices are limited to Unit-Disk graphs. In this setting, the inherent locality of the graphs can be leveraged by classical polynomial-time approximation schemes (PTAS) that guarantee an $\epsilon$-approximate solution. In this work, we present a deterministic and polynomial-time construction to embed a large family of non-local graphs in 3D atomic arrays. This construction is a first crucial step towards tackling combinatorial tasks on quantum computers for which no classical efficient $\varepsilon$-approximation scheme exists.
\end{abstract}

\maketitle
\section*{Introduction}



In the past decade, Quantum Processing Units (QPUs) consisting of atoms trapped in arrays of optical tweezers have been used extensively to address quantum simulation problems with spin systems\,\cite{Browaeys2020,henriet_quantum_2020}. The techniques and methods developed can also be leveraged to explore the resolution of combinatorial optimization problems, as all of Karp’s 21 NP-complete problems can be reformulated as ground-states of Ising models\,\cite{Lucas_2014}. The authors of the pioneer publication \,\cite{Pichler2018} noticed that the Hamiltonian of interacting Rydberg atoms naturally realizes the cost function of the Maximum Independent Set (MIS) problem on the graph $G$ induced by the atoms in interaction. This feature enables the implementation of adiabatic or variational schemes such as the QAA\,\cite{Farhi_adiabatic} or the QAOA\,\cite{Farhi2014} algorithms to approximately solve the MIS problem. The set of graphs that can be solved corresponds to Unit Disk (UD) graphs, where vertices are represented as points in the Euclidean plane and two vertices are connected by an edge if the distance between the two corresponding points is lower than a threshold value.\\

A recent implementation on more than 280 atoms has attracted a lot of attention\,\cite{Ebadi2022}, with the observation of heuristic resolution of the MIS problem on a large set of UD graphs. To grasp the performances of such NISQ algorithms, we need to compare them to their approximate classical counterparts. In the case of the MIS problem on UD graphs, there exist efficient classical approximation algorithms with guaranteed performance ratios (PTAS), which leaves little room for quantum advantage perspectives.\\

In order to leave the classes of efficient classical approximation schemes, we present a systematic and efficient method to map a MIS of any general graph of max-degree $6$ onto the ground state of an ensemble of interacting neutral atoms in 3D. For such graphs, no classical algorithm is known to find an $\varepsilon-$approximate solution in polynomial time. \\

The manuscript is structured as follows. In Section\,\ref{sec:MIS}, we recall some details about the MIS problem, and summarize how this problem can be solved with arrays of individual atoms in optical tweezers. In Section\,\ref{sec:UD-ptas}, we analyze the effects of graph locality on the approximability of the solutions to the problem. The main result of this paper is described in Sec. \ref{sec:embedding}, where we describe how to embed non-local graphs into 3D arrays of atoms with the addition of ancillas.\\

\section{Solving the MIS problem on local graphs\label{sec:MIS}}

Given a graph $G=(V, E)$, an independent set is defined as a subset $S$ of the nodes such that no two nodes of $S$ share an edge in $G$. Mathematically, $S$ is an independent set of $G$ \textit{iff} $S \subseteq V / \forall (x,y) \in S^2 , (x,y) \notin E$. A maximum independent set $S^*$ corresponds to an independent set of maximum cardinality. 

Any possible solution to this problem consists in separating the nodes of $G$ into two distinct classes: an independent one and the others. We attribute a status $z$ to each node, where $z_i = 1$  if node $i$ belongs to the independent set, and $z_i=0$ otherwise. The Maximum Independent Sets correspond to the minima of the following cost function: 

\begin{equation}
   C(z_1,\dots,z_N) = -\sum_{i=1}^N z_i + U \sum_{\langle i,j \rangle}z_i z_j
 \label{cost_function}
\end{equation}
where $U \gg \Delta(G)$, where $\Delta(G)$ is the degree of the vertex with  maximum degree, $\langle i,j \rangle$ represents adjacent nodes, and $N=|V|$. This cost function favours having a maximal number of atoms in the $1$ state, but the fact that $U \gg 1$  strongly penalizes two adjacent vertices in state 1. \\

Interestingly, the cost function of Eq. (\ref{cost_function}) can be natively realized on a neutral atom platform\,\cite{Pichler2018}, with some constraints on the graph edges. Placing $N$ atoms at positions $\textbf{r}_j$ in a 2D plane, and coupling the ground state $|0\rangle$ to the Rydberg state $|1\rangle$ with a laser system enables the realization of the Hamiltonian :
\begin{equation}
    H= \sum_{i=1}^N \frac{\hbar\Omega}{2} \sigma_i^x - \sum_{i=1}^N \frac{\hbar \delta}{2}  \sigma_i^z+\sum_{j<i}\frac{C_6}{|\textbf{r}_i-\textbf{r}_j|^{6}} n_i n_j.
\label{eq:ising_Hamiltonian}
\end{equation}
Here, $\Omega$ and $\delta$ are respectively the Rabi frequency and detuning of the laser system and $\hbar$ is the reduced Planck constant. The first two terms of Eq. (\ref{eq:ising_Hamiltonian}) govern the transition between states $|0\rangle$ and $|1 \rangle$ induced by the laser, while the third term represents the repulsive Van der Waals interaction between atoms in the $|1\rangle$ state. More precisely, $n_i = (\sigma_i
^z + 1)/2$ counts the number of Rydberg excitations at position $i$. The interaction strength between two atoms decays as $|\textbf{r}_i-\textbf{r}_j|^{-6}$ and $C_6$ is a constant which depends on the chosen Rydberg level. \\

The shift in energy originating from the presence of two nearby excited atoms induces the so-called \textit{Rydberg blockade} phenomenon. More precisely, if two atoms are separated by a distance smaller than the Rydberg blockade radius $r_b = (C_6/\hbar \Omega)^{1/6}$, the repulsive interaction will prevent them from being excited at the same time. On the other hand, the sharp decay of the interaction allows us to neglect this interaction term for atoms distant of more than $r_b$. As such, for $\Omega=0$, the Hamiltonian in Eq. (\ref{eq:ising_Hamiltonian}) is diagonal in the computational basis and enables to realize $H |z_1,\dots,z_N\rangle=(\hbar \delta/2) C(z_1,\dots,z_N)|z_1,\dots,z_N\rangle$, with the cost function specified in Eq. (\ref{cost_function}), and for which there is a link between atoms $i$ and $j$ if they are closer than $r_b$ apart.


\section{Local geometric structure imply PTAS}
\label{sec:UD-ptas}

\begin{figure}[ht]
    \centering
    \includegraphics[width =1.1\linewidth]{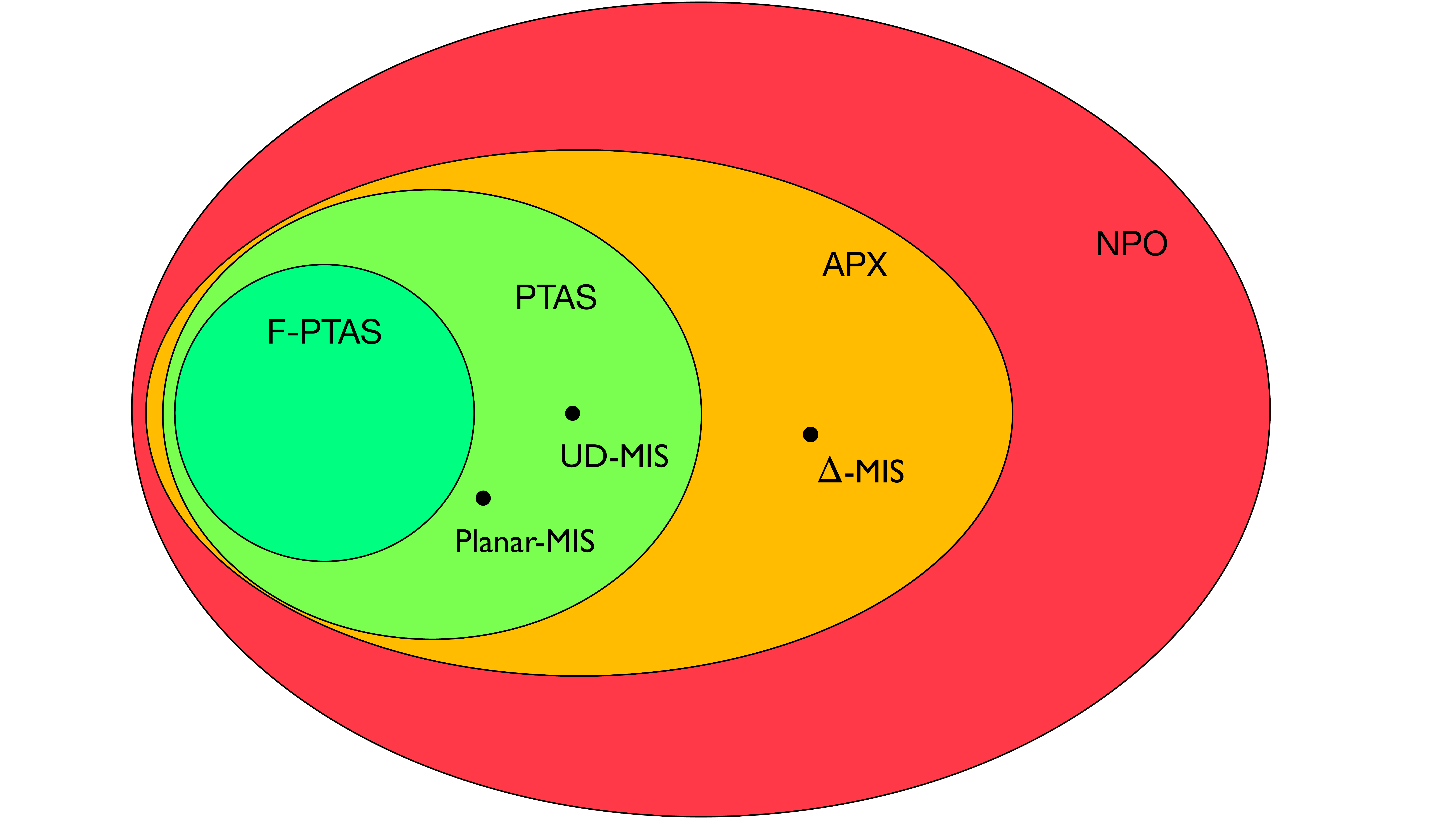}
    \caption{\textbf{Standard approximability classes of NP-hard problems} (under the assumption $P \neq NP$). While solving MIS exactly is NP-Hard, finding an approximate solution of useful quality can be easy, depending on the nature of the graph at hand. The complexity class NPO corresponds to the class of optimisation problems whose underlying decision problem is in NP. A polynomial-time approximation scheme (PTAS) is a family of $\varepsilon-$parameterised algorithms that output approximate optimal solutions that are $\varepsilon > 0$ away from the exact solution in polynomial time. When the family of algorithms is also polynomial in the parameter $\epsilon$, it is a Full-PTAS (F-PTAS) and is very efficient. The APX class corresponds to problems for which a polynomial-time algorithm can only achieve a constant approximation ratio. For MIS, if the graph is Unit-Disk or planar, there exists a PTAS. In our work, we propose a polynomial embedding of general bounded-MIS ($\Delta$-MIS) problems which are known to be in APX. $\Delta$-MIS is even APX-Complete, meaning that no PTAS exists for it unless $P = NP$.}
    \label{fig:Approx_classes}
\end{figure}

Unit-Disk graphs are inherently local in the sense that two nodes $v$ and $w$ are connected by an edge if and only if the distance between the two nodes is inferior to a given threshold. While finding the exact solution remains $NP-$hard, there exists efficient approximations to the solution. Known results about approximations to the general MIS problem are presented in Figure \ref{fig:Approx_classes}. Interestingly, the quality of the approximation depends on the type of graph under study: for Unit-Disk graphs, classical algorithms can leverage the locality of the edges to efficiently estimate an approximation of the MIS. The main idea is to split the graph into local subgraphs for which MIS is solved exactly. Aggregating the solutions of the subgraphs yields a good solution as a subgraph only affects its neighbouring subgraphs. It was shown that this method corresponds to a Polynomial-Time Approximation Scheme (PTAS) that guarantees a $1-\varepsilon$ approximation ratio in polynomial time\,\cite{hunt1998nc}. In the case of planar graphs, another PTAS exists that also guarantees high approximation ratios. In the same flavour as for the Unit-Disk case, it relies on dividing the graph into subgraphs with $k-$outerplanar forms \,\cite{Baker1994}.

\begin{figure*}[t!]
    \centering
    \includegraphics[width =0.8\linewidth]{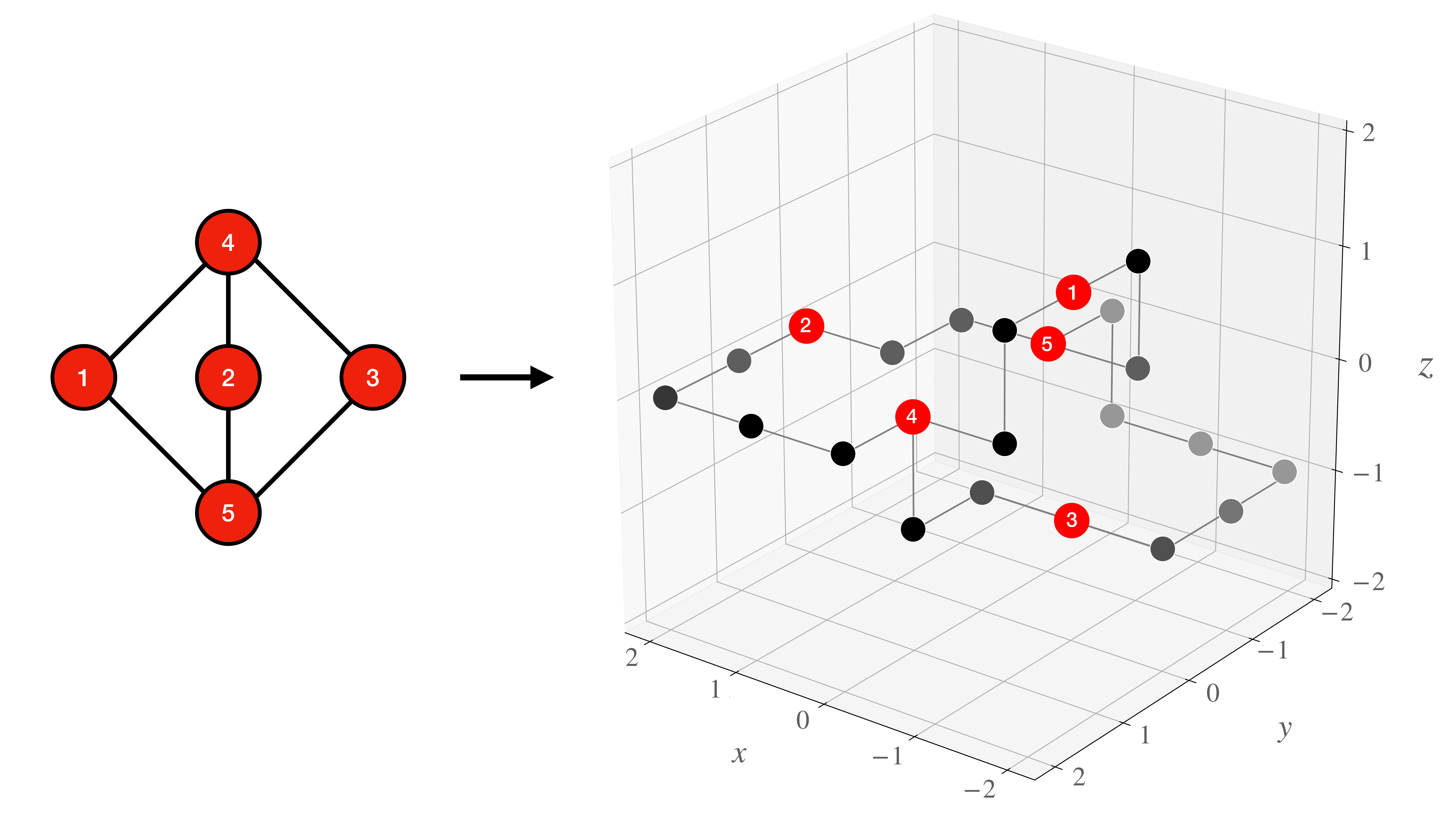}
    \caption{\textbf{Mapping a non-UD graph with Rydberg atoms using 3D quantum wires.}  The graph on the left is the smallest example of a non-Unit-Disk graph (otherwise the node 2 would be connected to 1 and 3). This graph is embedded in a 3D array of atoms (right). The red atoms correspond to the original graph nodes, while the chains of black atoms encode the edges of the original graph. The procedure enables to map any general graph, including non-UD and non-planar graphs with a bounded-degree up to $\Delta=6$.}
    \label{fig:3D_mapping}
\end{figure*}

However, more general graphs such as bounded-degree graphs do not present enough structure for classical algorithms to $\varepsilon-$approximate maximum independent sets in polynomial time. In the case of a graph with bounded-degree $\Delta$, finding an approximation solution to the MIS problem is known to be APX-complete\,\cite{Papadimitriou1991}. In other words, this means that the best approximation ratio that can be achieved by a classical algorithm is constant; to the best of our knowledge this approximation ratio is $r = \frac{5}{\Delta+3}$\,\cite{Berman1994}. This ratio cannot be improved without adding an exponential time overhead for a classical algorithm. A summary of the approximability classes of MIS on these specific classes of graphs is shown in Fig. \ref{fig:Approx_classes}.\\

The differences in the approximability of these problems motivate the need for efficient hardware embedding of graphs which present less geometrical structure than UD or planar graphs.


\section{Embedding bounded degree graphs in 3D arrays of atoms}
\label{sec:embedding}


Recent and previous works propose to represent non-local edges of graphs with chains of ancillary atoms\,\cite{Pichler2018, Ahn2021}, in 2D and 3D respectively. Building upon this idea, we present an efficient and automatic method to represent any graph of degree inferior or equal to 6 with a 3D array of atoms. An illustration of our method is given in Figure \ref{fig:3D_mapping}. It runs in polynomial time and numerical simulations suggest a low overhead in the number of added ancillary atoms (linear). 

\subsection{3D embedding}

We define the \textit{drawing} of a graph as the realisation or layout of a graph in a 3D space, where no two vertices overlap and no vertex-edge intersection occurs unless its incidence exists in the original graph. We say that a drawing is \textit{crossing-free} if no two edges cross. A growing interest emerged in 3D drawings of graphs for circuit designs\,\cite{leighton1986three} or for information visualisation\,\cite{ware1994viewing, Ware08}. In our case, we focus on 3D orthogonal grid drawings (OGD) of a graph $G=(V,E)$ for which the vertices of $G$ are represented as distinct points of the grid $\mathbb{Z}^3$, while all edges $E={(u,v) \in V^2}$ are restricted to being drawn on lines parallel to one of the three axes. This restriction implies that only graphs with maximum degree six can have such a drawing; given a vertex at coordinates $(x,y,z) \in \mathbb{Z}^3$, the only authorised directions for an edge are $(x \pm 1,y \pm 1,z \pm 1)$. It is proven that every graph of bounded-degree $\Delta$ admits a crossing-free OGD if $\Delta \leq 6$\,\cite{Barzdin1993, Eades1996}.\\

The general idea behind our method is to construct an OGD for the original graph and replace the edges by chains of ancillary atoms. The advantage of an OGD is that two distinct edges only intersect at common endpoints, thereby preventing ancillary atoms from interacting if they are not part of the same chain. Ideally, we would like to find an OGD that minimises the edge lengths in order to have as little ancillary atoms as possible. Unsurprisingly, it is $NP$-hard to find a OGD that minimises the total length of the edges\,\cite{Eades1996}. While many different algorithms have been proposed to optimise the total volume\,\cite{Biedl02orthogonaldrawings} or the average number of bends per edge\,\cite{Papakostas97} of an OGD, we present a simple and efficient method to construct an OGD with a small total length of edges.

Given a general $\Delta$-bounded graph $G=(V, E)$ with $\Delta \leq 6$, we place the nodes of $V$ in $\mathbb{R}^3$ using the Fruchterman-Reingold algorithm (FR)\,\cite{fruchterman1991graph} that runs efficiently in $\bigO(|V|^3)$ steps. Note that other algorithms could be used at this step but FR yielded the best results in our simulations. The nodes are then moved to the closest grid point in $\mathbb{Z}^3$, insuring that no two nodes get the same coordinates.

\begin{figure}
    \centering
    \includegraphics[width=0.8\linewidth]{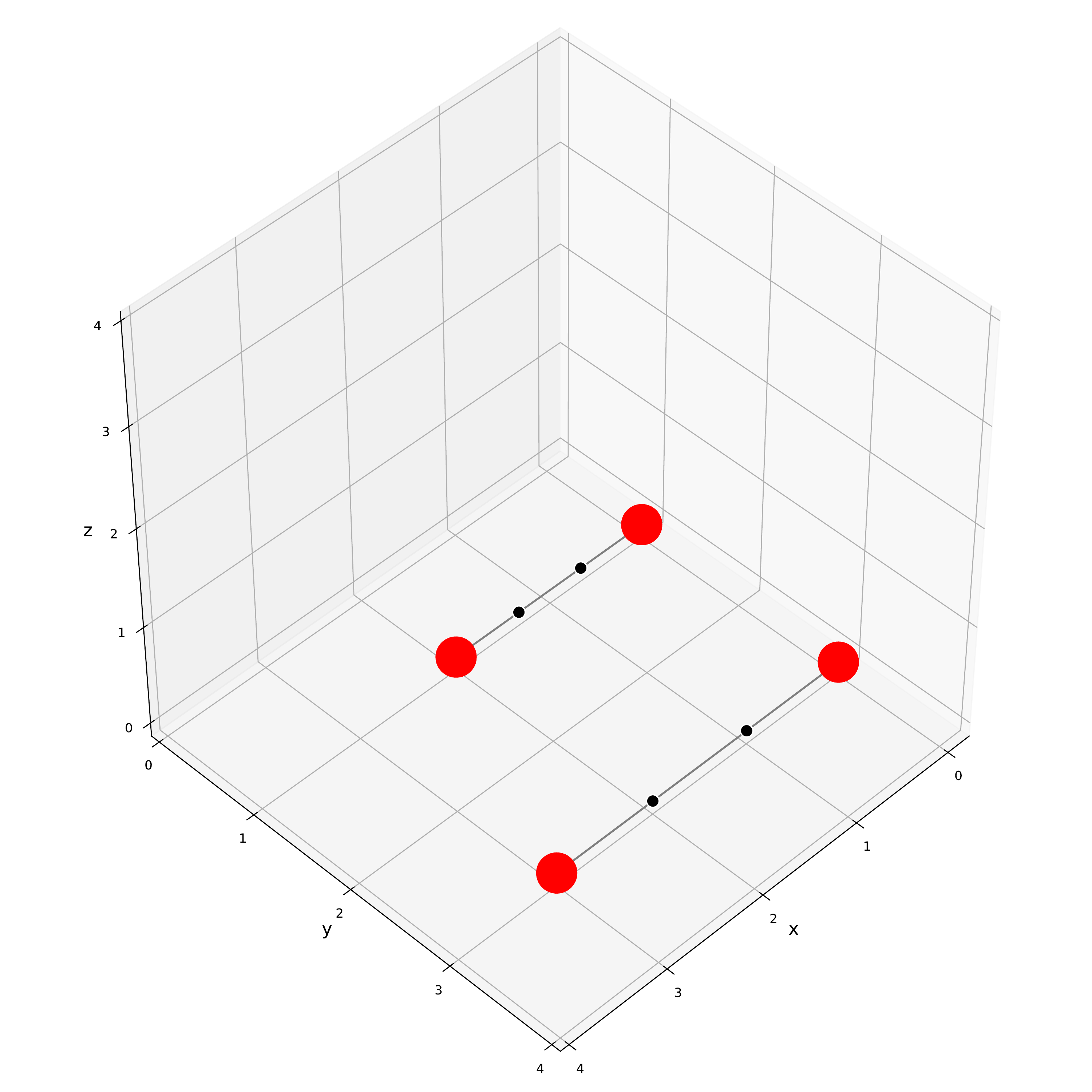}
    \caption{\textbf{Replacing edges with chains of ancillary atoms.} If the path length is odd, we add an ancillary node at each $\mathbb{Z}^3$ coordinate of the path. If the path is of even length $p$, we add $p$ evenly spaced ancillary vertices along the path. An even number of ancillary atoms per path is necessary to ensure that the MIS problem is rightly encoded in the augmented ensemble of atoms.}
    \label{fig:odd_line}
\end{figure}

We then use optimal path-finding algorithms\,\cite{Dijkstra1959,Hart1968} to find the shortest route between two nodes, restricted to the underlying grid. The resulting path is transformed into a chain of ancillary nodes. We previously ensure that two distinct edges are separated by at least a 2-grid-point distance, in order to avoid any unwanted interaction between ancillary atoms of two distinct chains. Finally, if the path length is odd, we add an ancillary node at each $\mathbb{Z}^3$ coordinate of the path. If the path is of even length $p$, we add $p+1$ evenly spaced ancillary vertices along the path as illustrated in Figure \ref{fig:odd_line}. After this procedure, one obtains an augmented graph $G_+=(V_+, E_+)$ of size $|V_+| = N_+$. 

\subsection{Encoding \texorpdfstring{$\Delta$}{Delta}-MIS in an augmented graph}

By representing all nodes with atoms, one can encode a Maximum Independent Set of $G$ in the ground state of an Ising Hamiltonian on Rydberg atoms\,\cite{Browaeys16} over the augmented graph $G_+$ :

\begin{equation}
H_+=\sum_{j=1}^{N_+}\frac{\hbar\Omega}{2}  \sigma_j^x- \sum_{j=1}^{N_+} \frac{\hbar}{2}(\delta + \delta_j)\sigma_j^z+\sum_{j<i}\frac{C_6}{|\textbf{r}_i-\textbf{r}_j|^{6}} n_i n_j.
\label{ising}
\end{equation} where $\delta_j$ represents the local detuning applied to each atom and $n_i=(\sigma^z_i + \mathbb{1})/2$.

\begin{figure}
    \centering
    \includegraphics[width=\linewidth]{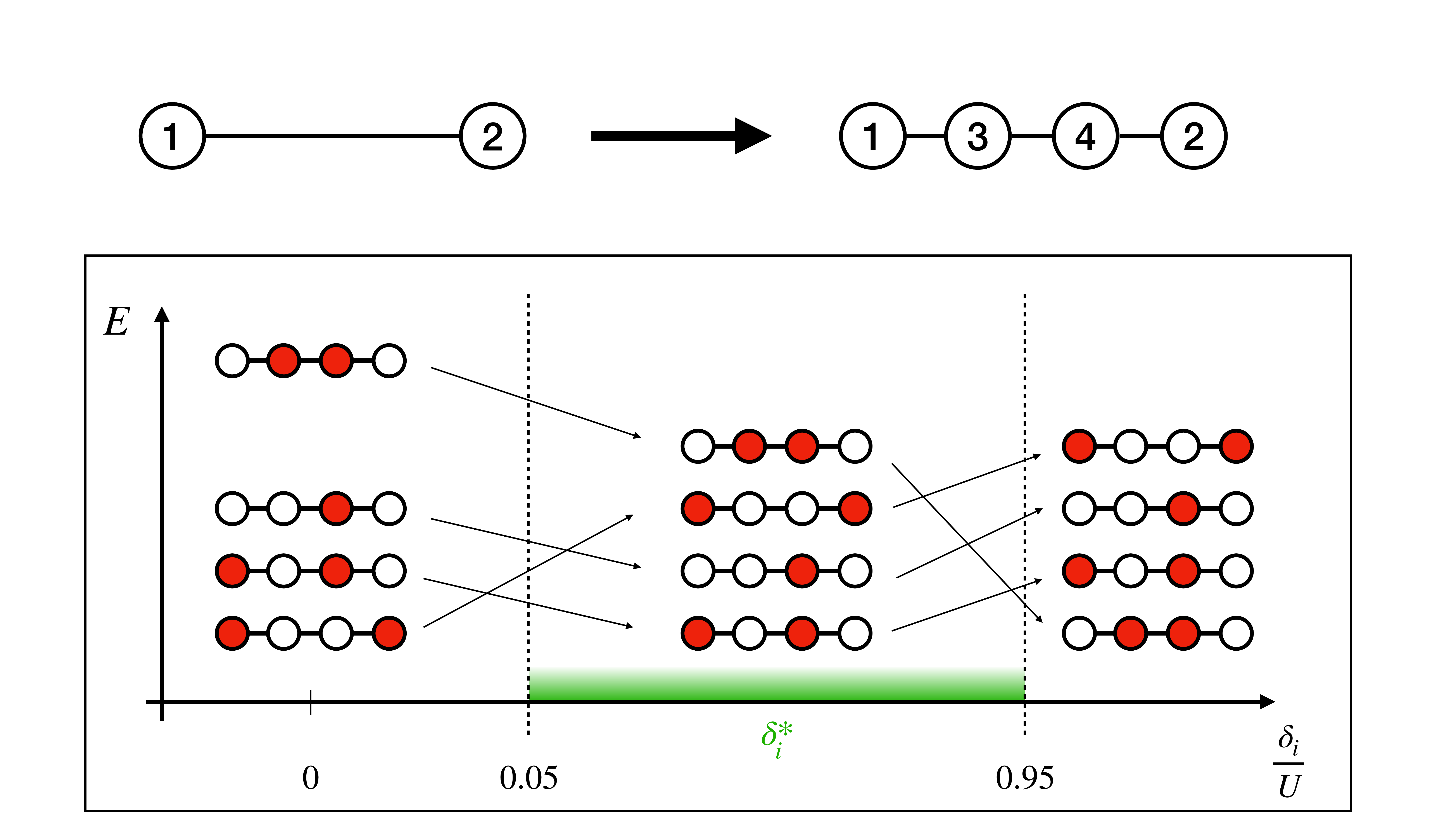}
    \caption{\textbf{Local detuning on ancillary atoms.} In the top, an edge between node 1 and 2 is augmented with two ancillary atoms 3 and 4. With no local detuning, the ground-state of the augmented edge corresponds to exciting $\{1,2\}$. This however is inconsistent in the initial graph (1 and 2 cannot be simultaneously in the MIS). We therefore add local detuning to all ancillary atoms to ensure that the ground-state corresponds to the anti-ferromagnetic state.  (b) Energy diagram of the spectrum as a function of the local detuning $\delta_i$ applied to the ancillary atoms. A red atom corresponds to an excited state. The independence condition of the original link is respected if $0.05 \times U \leq \delta_i \leq 0.95 \times U$, where $U$ is the interaction between two neighbouring atoms. In these values of local detuning, the ground-state corresponds to the anti-ferromagnetic state.}
    \label{fig:Augmented_path}
\end{figure}

In Figure \ref{fig:Augmented_path}, we show explicitly on a single augmented edge how one can choose the values of local detuning on ancillary atoms to ensure that the MIS of the edge corresponds to the groundstate of the Hamiltonian $H_+$.

In this example, there are 3 Maximum Independent Sets in the augmented graph : $\{1,4\},\{2,3\}$ which are acceptable solutions, but also $\{1,2\}$ which does not correspond to a MIS of the original graph. With a reasonable global detuning $\delta>0$, this latter state is actually the ground state of the chain. In order to guarantee that $\{1,4\}, \{2,3\}$ correspond to the ground-states of the Hamiltonian associated to the augmented path, we apply an additional local detuning $\delta_i$ to each ancillary atom with $\delta_i = J/2$, where $U=C_6/r^6$ is the interaction energy between two closest atoms of the augmented graph.

Keeping $0.5 \times U \leq \delta_i \leq 0.95 \times U$ ensures that the MIS returned by the algorithm preserves initial constraints (the proof is given in appendix \ref{appendix_local_detuning}). In the general case, we determine for each edge the corresponding value for the detuning that will be applied on all the ancillary atoms of that chain. This ensures that the ground state of the augmented Hamiltonian encodes a Maximum Independent Set of the original graph.

\subsection{Overhead}
  
\begin{figure}[ht]
    \centering
    \includegraphics[width =1.\linewidth]{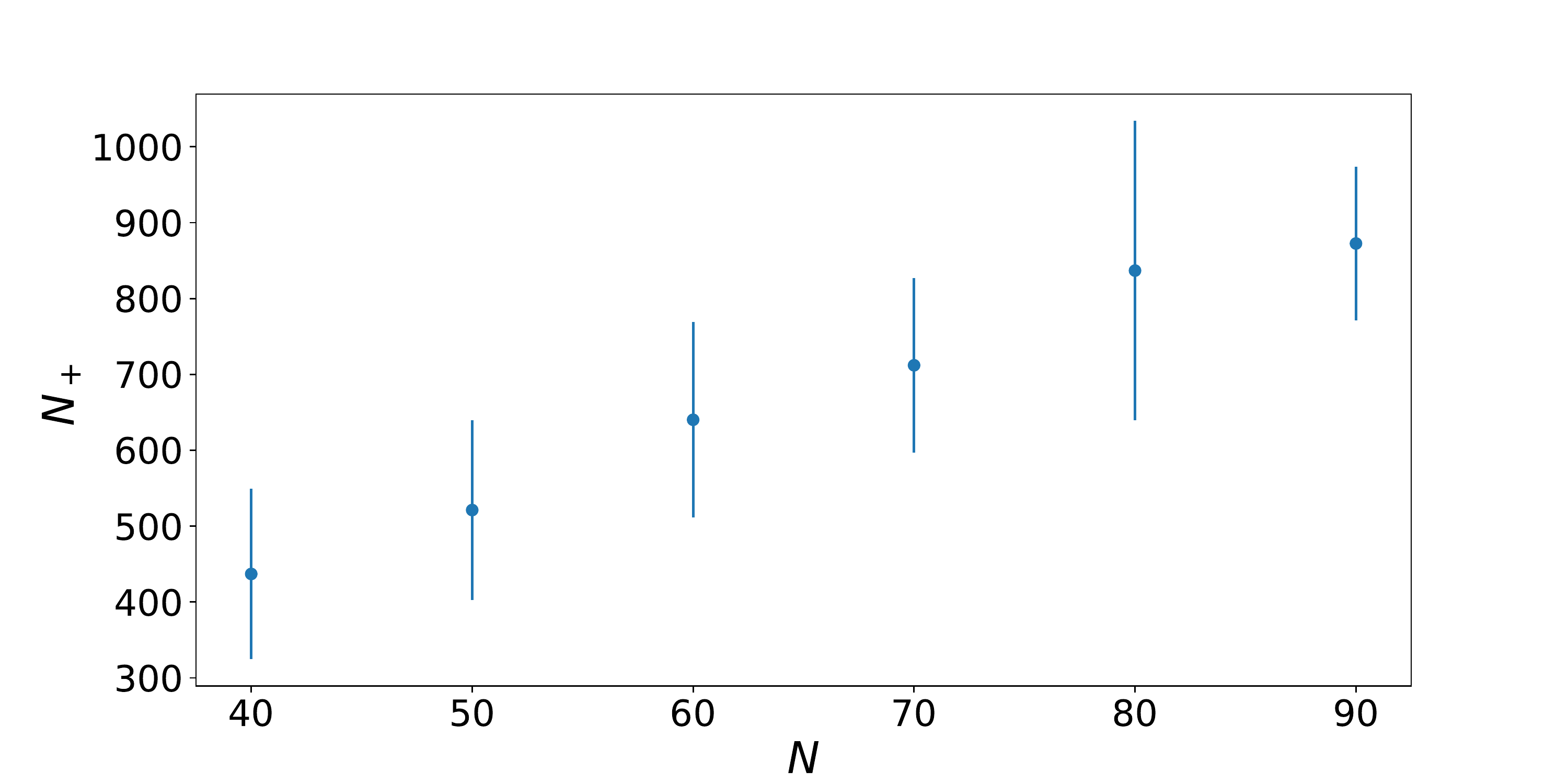}
    \caption{\textbf{Scaling of additional atoms in our method}. For each size, 20 random Erdos-Renyi graphs of maximum degree 6 are generated. Our method is then applied and we plot the mean size $N_+$ of the augmented graphs and the standard deviation of the sample.}
    \label{fig:overhead_ancillary}
\end{figure}

The procedure described above enables us to replicate the connectivity of any input graph G of maximum degree six in three dimensions, at the expense of adding ancillary nodes. In order to assess the overhead incurred by this embedding, we test the procedure of general graphs of maximum degree 6. For each size, 20 Erdos-Renyi graphs are generated and the size $N_+$ of the augmented graph is recorded. Our simulations seem to indicate a linear overhead in the number of ancillary atoms, as illustrated in Figure\,\ref{fig:overhead_ancillary} where we show the number of nodes in the augmented graph $N_+$ with respect to the size of the original graph $N$. Our method would be impractical if the number of ancillary atoms exploded or if the size of the edges grew exponentially with respect to the graph size. Indeed, the Lieb-Robinson bounds\,\cite{lieb1972finite} would imply that information could not propagate efficiently through the ancillary paths. Luckily, the necessary volume to draw the OGD of a graph was proven to be polynomially bounded\,\cite{Eades1996}. Precisely, every $N$-vertex degree-6 graph admits an OGD in $\mathcal{O}(N^{3/2})$ volume and that bound is best possible for some degree-6 graphs. The authors give an explicit algorithm that places all vertices on a $\mathcal{O}(N) \times \mathcal{O}(N)$ grid in a $2D$ plane and draws each edge with at most $16$ bends. The growing number of atoms that can be experimentally trapped in recent experimental setups\,\cite{Schymik_2022} is a encouraging sign that the overhead in the number of atoms required in our method is reasonable.

\section*{Conclusion}

We presented a method to embed the MIS problem over non-local graphs using a 3D array of atoms trapped in optical tweezers. This procedure is guaranteed to run in a time growing polynomially with the input graph size. The main trick consists in representing non-local edges by chains of ancillary atoms and adding local detuning to them in order to ensure that the ground-state of the system encodes the solution to the non-local MIS. One could then use quantum annealing or QAOA to prepare low-energy states of the system. An important aspect is that the class of graphs that can be tackled with our method corresponds to bounded-degree graphs of maximum degree 6, for which no PTAS exists unless $P = NP$. A very intriguing and exciting prospect is to qualify both theoretically and experimentally the capabilities of quantum devices in approximately solving the MIS on this class of graphs. We believe that understanding from a theoretical perspective the capabilities of quantum approaches in finding guaranteed performance ratio  for NP-Complete problems is an important question in the quest for practical quantum advantage. 

\section*{Acknowledgments}
We thank March Porcheron, Lucas Leclerc, Alex B. Grilo, Elham Kashefi, Jaewook Ahn, Thierry Lahaye and Antoine Browaeys for thoughtful discussions and remarks. We acknowledge support from the region Ile-de-France through the AQUARE project, as part of the PAQ program. After completion of this manuscript, we became aware of a related work\,\cite{Nguyen22} aimed at encoding graphs beyond hardware constraints with Rydberg atoms arrays.

\appendix
\section{Optimal local detuning on ancillary atoms}
\label{appendix_local_detuning}

We estimate in this proof the lower and upper bound for the local detuning on the two ancillary atoms of an augmented edge. Let $E_{i_1i_3i_4i_2}$ be the energy associated to the bit-string $i_1i_3i_4i_2$, where $i_k \in \{0,1\}$ and $k$ is the label of the atom ($1,2$ are the main atoms, and $3, 4$ the ancillas. We want to ensure the following inequalities: 
\begin{equation}
    \begin{cases} 
    E_{1010} < E_{1001} \\ 
    E_{1010} < E_{0110} \\
    \end{cases}
\end{equation}

We show calculations in the case $\delta_i = 0$ for original atoms (1 and 2). We therefore have that 

\begin{equation}
\begin{split}
    &\begin{cases} 
    -2\delta - \delta_i + \frac{1}{2^6}U < -2\delta + \frac{1}{3^6}U  \\ 
    -2\delta - \delta_i + \frac{1}{2^6}U  < -2 (\delta + \delta_i) + U \\
    \end{cases} \\
    &\iff \delta_i \in [( \frac{1}{2^6} - \frac{1}{3^6}) \times U, (1 - \frac{1}{2^6}) \times U]
\end{split}
\end{equation}

Taking $\delta_i = U/2$ for all ancillary atoms is a safe spot.
\bibliographystyle{unsrtnat}
\bibliography{refs}

\end{document}